\documentclass[useAMS,usenatbib]{mn2e}
\usepackage{times}
\usepackage{graphicx}
\usepackage{amssymb}
\usepackage[mathscr]{eucal}
\usepackage{verbatim}
\usepackage{textcomp}

\title{Angular momentum in cluster Spherical Collapse Model}
\author[Guido Cupani, Marino Mezzetti, Fabio Mardirossian]
{Guido Cupani$^1$\thanks{E-mail: \texttt{cupani@oats.inaf.it}.}
Marino Mezzetti$^{1,2}$
and Fabio Mardirossian$^{1,2}$\\
$^1$ INAF - Istituto Nazionale di Astrofisica, via Tiepolo 11, I-34143 Trieste, Italy\\
$^2$ Dipartimento di Astronomia, Universit\`a degli studi di Trieste, via Tiepolo 11, I-34143 Trieste, Italy
}
\pagerange{\pageref{firstpage}--\pageref{lastpage}}
\pubyear{2008}

\newcommand{\bi}{\begin{itemize}}
\newcommand{\be}{\begin{equation}}
\newcommand{\ei}{\end{itemize}}
\newcommand{\ee}{\end{equation}}

\newcommand{\f}{\frac}
\newcommand{\p}{\item}

\begin{document}

\maketitle

\begin{abstract}
Our new formulation of the Spherical Collapse Model (SCM-L) takes into account the presence of angular momentum associated with the motion of galaxy groups infalling towards the centre of galaxy clusters. The angular momentum is responsible for an additional term in the dynamical equation which is useful to describe the evolution of the clusters in the non-equilibrium region which is investigated in the present paper. Our SCM-L can be used to predict the profiles of several strategic dynamical quantities as the radial and tangential velocities of member galaxies, and the total cluster mass. 

A good understanding of the non-equilibrium region is important since it is the natural scenario where to study the infall in galaxy clusters and the accretion phenomena present in these objects. Our results corroborate previous estimates and are in very good agreement with the analysis of recent observations and of simulated clusters. 
\end{abstract}


{\setlength\arraycolsep{1pt}

\section{Introduction}\label{sec:intro}


The cosmic structures we observe in the Universe are thought to be originated by small perturbations in the primordial  matter distribution, enhanced in time by self-gravity. Density perturbations decouple from the Hubble flow and drive the collapse of matter into structures (groups and clusters of galaxies) through a highly non-linear process. In a Universe with cold dark matter and a cosmological constant ($\Lambda$-CDM), structures of smaller scale decouple before structures of larger scales do: this means that galaxy clusters are built up through merging and coalescence of smaller galaxy groups. This ``bottom-up'' scenario has been widely tested using large $N$-body hydrodinamical simulations and is consistent with the most recent measurement of the cosmological parameters.

An accurate general description of the cosmic structure collapse is analitically inpraticable. To tackle the problem, simplifying assumptions are required. The most common and accepted assumption is to consider only spherically symmetric density perturbations: in this case, each perturbation can be described as an independent Friedmann universes on its own, with constant positive curvature, which first expands and then collapses mantaining its spherical shape. This Spherical Collapse Model (SCM) was first developed by \cite{GG}, \cite{S74}, and \cite{G}, and it has since became a widely accepted approach in the literature (e.g.~\citealt{P76,P80,Sch,Shm,Oal}). Many authors demonstrated that despite its simplistic assumption the SCM provides a quite good description of the dynamics in the outskirts of clusters, where the matter is not in energy equilibrium and is characterized by an overall infall motion towards the cluster centre (e.g.~\citealt{Pal,BRal,BC,CMM08}, hereafter CMM08). Corrections to the SCM are generally required in the cluster cores, due to the interpenetration of different collapsing shells of matter (an effect referred to as shell crossing; see e.g.~\citealt{SC} and references therein).

One of the major shortcomings of the SCM is that in principle a purely spherically symmetric scenario does not allow for any non-radial motion of the collapsing matter. In fact, galaxy dynamics in observed clusters show a significant amount of tangential motion \citep{RG}. \cite{DG} demonstrated that the amplitude of the caustic surfaces observed in the redshift-space distribution of galaxies is related to the galaxy escape velocities, which are by definition non-radial (see e.g. \citealt{D99,Ral00,Ral01,Ral03,RD}; and \citealt{D09} for a review). More recently, \cite{CMM10} showed that radial velocities and tangential velocities of galaxies are statistically of the same order of magnitude in modulus. Observation of line-of-sight velocities can therefore be used to estimate the infall pattern and the mass profile of clusters up to the outermost regions (corresponding to radii exceeding seven times the virialization radius).

In this paper we take into account the effect of tangential motion into the SCM. Considerations based on the $\Lambda$-CDM bottom-up scenario suggest that angular momentum in cluster outskirts is mainly associated with galaxy groups which belongs to the clusters, rather than single galaxies. In fact, galaxies rarely infall isolated, since groups appear to be formed long before they merge into clusters. We will show that the tangential motion of infalling galaxies is not only due to the intrinsic velocity dispersion of the infalling groups which the galaxies belong to, but also to the coupled effects of the angular momentum conservation and the torque produced by the presence of nearby structures. Several authors discussed the effect of the angular momentum in cosmic structures, but they mainly focused on the growth of the total angular momentum and/or on the behaviour of the inner, relaxed regions (e.g.~\citealt{RG,R88,CT,ARal,N,Aal,DPK}). On the contrary, we will not take into account here the overall angular momentum of the galaxy cluster which can arise from the misalignment between the inertia tensor and the tidal forces acting on the collapsing structures \citep{W}, nor the process of violent relaxation \citep{LB}, which transforms the infall energy into non-radial velocity dispersion on a collapse timescale. But we will follow a different approach and focus on the non-equilibrium region of clusters, where the presence of angular momentum is expected to slow down the overall infall motion of matter. We will show that the SCM with angular momentum (hereafter SCM-L) better describes, compared to the standard SCM, the infall velocity profile and the mass profile in the outskirts of cluster; in fact it predicts shallower infall velocity profiles which are consistent with the results of CMM08 about the velocity profiles. 

We wish to stress that considering the effects of the angular momentum in the dynamics of the outskirts of clusters necessarily implies an approach which is linked to the dynamics of the separate structures infalling towards the cluster core, so we have not to deal with the cluster total angular momentum.

In the following, we will discuss how to to take into account the effect of the angular momentum and how to derive it from the power spectrum of the density perturbations which originated the clusters (section \ref{sec:proc}). The derived  equation provides information about both the matter distribution and the kinematical properties of clusters. We will thoroughly compare these predictions with our results published in CMM08 and CMM10 (section \ref{sec:results}) and discuss their implications in cosmology (section \ref{sec:concl}). The details of the calculation are generally omitted in the main text, and a full account on them can be found in the appendices.

In the present paper we adopt $H_0=70$ km s$^{-1}$ Mpc$^{-1}$.
\section{Inserting angular momentum}\label{sec:ang_mom}

In the following, we detail the SCM-L scenario. We consider a spherical region with centre O, surrounded by an (almost) isotropic Universe with its own fluctuations. We assume at first that this region is actually completely homogeneous, i.e. there is not a cluster in it. Let us consider an object (within a mass shell) at a distance $a$ from  $\rmn O$, with a tangential velocity $v_\perp$. The evolution of this $v_\perp$ is described by the evolution of the specific (i.e. per unit mass) angular momentum $\mathcal L$ with respect to $\rmn O$:
\be\label{eq:L}
\f{\rmn d\mathcal L}{\rmn dt}=\f{\rmn d}{\rmn dt}(a v_\perp)=a\mathcal F_\perp,
\ee
where $a\mathcal F_\perp$ is the specific torque and $\mathcal F_\perp$ is the specific transversal force. 

Let us suppose now that in the same spherical region there is a galaxy cluster centered in $\rmn O$. In this case the mass shell, that was at a distance $a$ from the centre in the previous case, is now at a distance $r$ from $\rmn O$. The specific angular momentum $\bar\mathcal L$ of an object in the mass shell at the distance $r$ from the center and with transversal velocity $\bar v_\perp$ will evolve according to
\be\label{eq:barL}
\f{\rmn d\bar\mathcal L}{\rmn dt}=\f{\rmn d}{\rmn dt}(r\bar v_\perp)=r\bar\mathcal F_\perp,
\ee
where $\bar v_\perp$, $\bar\mathcal L$, and $\bar\mathcal F_\perp$ are barred to indicate that in principle they may be different with respect to the previous case.

The velocity of an infalling system is perturbed by the force produced by external structures. Structures on scales comparable to, or larger than the cluster scale will affect the overall motion of the cluster and of its outskirts, in such a way that the \emph{relative} velocity between the cluster and its outskirts will not change. On the other hand, structures on scales comparable to, or smaller than, the group scale are not big enough to significantly affect the infall motion of groups. So, only structures in a range of scales will affect the \emph{relative} velocities of the infalling systems. As we will later justify, the dominant contribution comes from scales on the order of $\sim 10 h^{-1}$ Mpc, and the difference between $r$ and $a$ are of the order of a Mpc, or less. We will therefore neglect the difference between $\mathcal F_\perp$ and $\bar\mathcal F_\perp$ inside the cluster, and we will assume they are equal, making the problems analytically tractable:
\be\label{eq:L_comp}
\f{\rmn d\bar\mathcal L}{\rmn dt}=\f{r}{a}\f{\rmn d\mathcal L}{\rmn dt}.
\ee

In order to estimate $\mathcal L=a v_\perp$ in the  \emph{no-cluster}  case, we can use the peculiar velocity $v$,  obtained through the continuity equation linking the density contrast $\delta$ to $v$ (see, e.g., \cite{Pea99}). The ortogonal velocity $v_\perp$ is obtained as the projection of $v$ onto a plane perpendicular to the radial direction. Under the isotropy assumption, this gives $v_\perp=\sqrt{2\langle v^2\rangle/3}$, while radially this gives a contribution $|{v_\parallel}|=\sqrt{\langle v^2\rangle/3}$  to the infall velocity. 

The knowledge of $\bar\mathcal L=r v_\perp$ allows to insert it into the  Friedman equation as follows: 
\be\label{eq:fried}
\f{\rmn d^2 r}{\rmn dt^2}=-\f{4\pi G\bar\rho}{3}r+\f{\Lambda c^2}{3}r+\f{\bar \mathcal L^2}{r^3}
\ee
where $\bar\rho$ is the mean density of the perturbation within $r$, $G$ is the gravitational constant, $\Lambda$ is the cosmological constant, and $c$ is the speed of light. Equation (\ref{eq:fried}) gives an exhaustive description of the SCM-L.

\section{Procedure}\label{sec:proc}

Three main statuses can be recognized in the evolution of a density perturbation:
\begin{enumerate} 
\p A primordial status, when the pertubation uncouples from the Hubble flow and expands at a slower rate than that of the unperturbed Universe; 
\p A turnaround status, when the perturbation reaches its maximum amplitude and starts collapsing;
\p A virialization status, when the collapse eventually ends and the dynamical equilibrium is reached.
\end{enumerate}
These statuses are reached at different times, and they depend on the primordial overdensity of the perturbation which originated the cluster. Overdensities which reach either the turnaround status or the virialization status at the end of their evolution are particularly interesting for studying the evolution of the clusters (see e.g.~\citealt{NFW95,NFW96,NFW97,Bal,LM}; ~\citealt{CMM08,CMM10}). The SCM allows us to analyze these two dynamical statuses stopping the cluster evolution at different times (which can or cannot coincide with the present time $t_0$). We use throughout the subscripts $\rmn t$ and $\rmn v$ to refer to the turnaround stage and to the virialization stage, and the subscripts $\rmn i$, $\rmn f$, and $0$ to refer to the initial time and the final time of the evolution, and to the present time, respectively. 

The cosmology is defined in agreement with the $\Lambda$-CDM prescription:
\be\label{eq:cosm}
\Omega_\rmn M\equiv\Omega_\rmn b(z)+\Omega_\rmn{DM}(z);\quad\Omega_\rmn M(z)+\Omega_\Lambda(z)=1,\quad\forall z; 
\ee
\be\label{eq:cosm_ev}
\Omega_\rmn M(z)=\f{\Omega_{\rmn M,0}(1+z)^3}{\Omega_{\Lambda,0}+\Omega_{\rmn M,0}(1+z)^3}.
\ee
Here $\Omega_\rmn X$ denote the density fraction of the species $X$ with respect to the critical density. The subscripts are defined as follows: $\rmn M$ = matter, $\rmn b$ = baryonic matter, $\rmn{DM}$ = dark matter, $\Lambda$ = dark energy or cosmological constant. We adopt throughout the concordance value for the baryonic density fraction, $\Omega_{\rmn b,0}=0.04$ (e.g., \citealt{Borgal}, \citealt{Borgal4}), and we also choose many other values for the matter density fraction, namely $\Omega_{\rmn{M},0}=0.20$, $0.25$, $0.30$, $0.35$, and $0.40$. We use $\Omega_{\rmn{M},0}$ values usually rejected by the literature, just in order to investigate the robustness of our results for different $\Omega_{\rmn{M},0}$.

The power spectrum of the perturbations $P(k)$ (with $k$ the perturbation wave number) is defined, according to the prescription of \cite{EH}, using their transfer function $T(k)$ (accurate to better than 5 per cent for $\Omega_{\rmn{b},0}/\Omega_{\rmn M,0}\la 0.5$): 
\be\label{eq:trans}
T(k)=\f{\Omega_{\rmn{DM},0}}{\Omega_{\rmn M,0}}T_\rmn{DM}(k)+\f{\Omega_{\rmn b,0}}{\Omega_{\rmn M,0}}T_\rmn b(k);\quad P(k)=kT(k)^2.
\ee
Here $T_\rmn b(t)$ is the component associated to the baryonic matter and $T_\rmn{DM}(k)$ is the component associated to the dark matter.  The resulting power spectrum is normalized to the root-mean-square density fluctuations within a sphere with radius $r_8\equiv 8h^{-1}\,\rmn{Mpc}$, $\sigma_8$, imposing, as in the reference simulation of  \citealt{Borgal4}, $\sigma_8=0.8$, in agreement with the value obtained by the latest measurements of the \emph{Wilkinson Microwave Anisotropy Probe} (WMAP) (see e.g.~\citealt{Kal}):
\begin{eqnarray}\label{eq:sigma_8}
\sigma_8&\equiv&\f{1}{2\pi}\int_0^\infty{k^2 P(k)\left[\f{3(\sin kr_8-kr_8\cos kr_8)}{(kr_8)^3}\right]^2\rmn dk}\nonumber\\
&=&0.81\pm 0.03.
\end{eqnarray}

Both the initial overdensity $\delta_\rmn i$ and the angular momentum $\bar\mathcal L$ of the perturbation are extracted from the power spectrum $P(k)$. To compute $\delta_\rmn i$ (within $r_\rmn i$), we apply a gaussian filter to the power spectrum, with radius $r_\rmn G=1h^{-1}\rmn{Mpc}$, since we are interested on progenitors at the galaxy cluster scale (see appendix \ref{app:delta_i}). To compute $\bar\mathcal L$, we use a band-pass filter, peaked at a scale of $\sim 10 h^{-1}$ Mpc, to isolate the contribution at the scale of galaxy groups (see appendix \ref{app:fried}). The last band pass filter is designed on the assumption that galaxy groups are mainly responsible for the angular momentum term which has to be introduced in the standard SCM. This assumption is consistent with the widely accepted bottom-up scenario (which looks at galaxy groups as the building blocks of galaxy clusters), and it will be justified \emph{a posteriori} by the dynamical analysis of the catalogue of simulated galaxy clusters we use (see section \ref{sec:prof}). The use of both the gaussian filter, both the band-pass filter, cuts away almost all the non-linear part of the power spectrum. For this reason, in the present paper, we used the linear $P(k)$.

The evolution of the perturbation is described by time integration of equations (\ref{eq:L_comp}) and (\ref{eq:fried}). We need as input $\delta_\rmn i$ and $\bar\mathcal L_\rmn i$ in order to obtain as output $R_\rmn f\equiv r_\rmn f/a_\rmn f$, $\left.\rmn dR/\rmn d\tau\right|_\rmn f$, $\Delta_\rmn f\equiv (1+\delta_\rmn i)R_\rmn f^{-3}$, and $\mathcal L_\rmn f$. The integration is performed between an initial time $\tau_\rmn i$ and the final time $\tau_\rmn f$, corresponding to initial redshift value $z_\rmn i=1000$ and the final value $z_\rmn f$, which depends on our choice of the end of the evolution, i.e. $z_\rmn f=0$, $0.1$, $0.2$, $0.3$, and $0.5$, which correspond to possible redshifts of observed clusters. $\tau$ is defined as
\be\label{eq:z_t}
\tau=\f{2}{3\sqrt{1-\Omega_{\rmn M,0}}}\rmn{asinh}\left[\sqrt{\f{1-\Omega_{\rmn M,0}}{\Omega_{\rmn M,0}}}(1+z)^{-3/2}\right],
\ee
according to the adopted cosmology.

We focused on the cases of a perturbation reaching either the turnaround state or the virialization state at the end of its evolution. The turnaround state is defined by the condition $\left.\rmn dR/\rmn d\tau\right|_\rmn f=0$; we used this condition to obtain the initial and final radius, $r_\rmn{i,t}$ and $r_\rmn{f,t}$, and the initial and final overdensity, $\delta_\rmn{i,t}$ and $\delta_\rmn{f,t}$, of a perturbation evolving into the turnaround status just at $\tau_\rmn f$. The virialization status is quite ambiguous, since neither the SCM nor the SCM-L are able to describe the details of final virialization, when shell crossing occurs. 
However, it is usually assumed in the literature that virialization occurs when, in the SCM scenario, the overdensity collapses into a point. In the SCM-L scenario, angular momentum prevents the collapse, and we have assumed for analogy that virialization corresponds to the state when
 $R_\rmn f$ reaches its first minimum.
We call $r_\rmn{i,v}$ and $\delta_\rmn{i,v}$ the initial radius and the initial overdensity of a perturbation which corresponds to this condition.

We computed the turnaround radius, $R_\rmn t\equiv r_\rmn{f,t}/a_\rmn f$, and the turnaround overdensity, $\Delta_\rmn t\equiv 1+\delta_\rmn{f,t}$, of this evolved perturbation. The ratio between turnaround mass and virialization mass is
\be
\f{M_\rmn t}{M_\rmn v}=\f{1+\delta_\rmn{i,t}}{1+\delta_\rmn{i,v}}\left(\f{r_\rmn{i,t}}{r_\rmn{i,v}}\right)^3,
\ee
while the ratio between turnaround radius and virialization radius is
\be
\f{r_\rmn t}{r_\rmn v}\equiv\left(\f{\Delta_\rmn v}{\Delta_\rmn{f,t}}\f{M_\rmn t}{M_\rmn v}\right)^{1/3},
\ee
where $\Delta_\rmn v\equiv[18\pi^2+82(\Omega_{\rmn M,0}-1)-39(\Omega_{\rmn M,0}-1)^2]/\Omega_{\rmn M,0}$ is the standard virialization overdensity (e.g., \citealt{BN}).

We follow the evolution of an overdensity profile from $\tau_\rmn i$ to $\tau_\rmn f$. We mapped the initial overdensity profile as a set of pairs $\{r_\rmn i,\delta_\rmn i\}_k$, with $k=1,\ldots,10$, each one describing a single perturbation, and let the pairs evolve into $\{R_\rmn f,\Delta_\rmn f\}_k$, to obtain the final overdensity profile. We are also able to compute the final profile of the radial velocity $v_{\parallel,\rmn f}$ and the tangential velocity $v_{\perp,\rmn f}$ (see appendix \ref{app:fried}). In the analysis, we focused on the following quantities, which are relevant for the dynamics on the non-equilibrium region of galaxy clusters (CMM10):
\be
\tilde M\equiv\f{M_\rmn f}{M_\rmn t}=\f{\Delta_\rmn f R_\rmn f^3}{\Delta_\rmn t R_\rmn t^3},
\ee
\be
\tilde v_\parallel\equiv\f{v_{\parallel,\rmn f}}{H_0 r_\rmn f}=\f{1}{\tilde R_\rmn f}\left.\f{\rmn d\tilde R}{\rmn d \tau}\right|_\rmn f,
\ee
\be
K_v\equiv-\f{v_{\perp,\rmn f}}{v_{\parallel,\rmn f}} 
\ee
where the tilde denotes normalization to turnaround values, as suggested by CMM08.


The results of our analysis are discussed in the next section. To assess the difference between the SCM-L and the standard SCM, we switched on and off the angular momentum term in the evolution equations (\ref{eq:L_comp}) and (\ref{eq:fried}). 


\section{Results}\label{sec:results}

In section \ref{sec:turn} we show that the SCM-L better estimates the turnaround overdensity, while no large differences are seen in the SCM and SCM-L ratios between the turnaround radius and the virialization radius, and in the SCM and SCM-L ratios between the turnaround mass and the virialization mass. 

In section \ref{sec:prof} we show that both SCM and SCM-L give quite good results in the determination of the cluster mass profiles, but only the SCM-L succeeds in a good fit of the radial velocity profile of the cluster components.

\subsection{Turnaround}\label{sec:turn}

\begin{table}
\caption{Ratios between the SCM and the SCM-L outcomes for $R_\rmn t$, i.e. $R_\rmn{t,SCM}/R_{\rmn{t,SCM}-\rmn{L}}$, for different choices of $\Omega_\rmn{M,0}$ and $z_\rmn f$.}
\label{tab:Rt}
\centering
\begin{tabular}{ccccccc}
&&&& $\Omega_{\rmn M,0}$ &&\vspace{0.1cm}\\
&& $0.20$ & $0.25$ & $0.30$ & $0.35$ & $0.40$\\
\hline
& $0.5$ & $0.85$ & $0.85$ & $0.85$ & $0.85$ & $0.85$\\
& $0.3$ & $0.83$ & $0.83$ & $0.83$ & $0.83$ & $0.84$\\
$z$\hspace{0.1cm} & $0.2$ & $0.83$ & $0.82$ & $0.82$ & $0.82$ & $0.82$\\
& $0.1$ & $0.82$ & $0.81$ & $0.81$ & $0.81$ & $0.81$\\
& $0$ & $0.81$ & $0.80$ & $0.80$ & $0.80$ & $0.80$\\
\hline
\end{tabular}
\end{table}

\begin{table}
\caption{Ratios between the SCM and the SCM-L outcomes for $\Delta_\rmn t$, i.e. $\Delta_\rmn{t,SCM}/\Delta_{\rmn{t,SCM}\textrm-\rmn{L}}$ for different choices of $\Omega_\rmn{M,0}$ and $z_\rmn f$.}
\label{tab:Deltat}
\centering
\begin{tabular}{ccccccc}
&&&& $\Omega_{\rmn M,0}$ &&\vspace{0.1cm}\\
&& $0.20$ & $0.25$ & $0.30$ & $0.35$ & $0.40$\\
\hline
& $0.5$ & $1.16$ & $1.16$ & $1.16$ & $1.15$ & $1.15$\\
& $0.3$ & $1.19$ & $1.20$ & $1.20$ & $1.20$ & $1.19$\\
$z$\hspace{0.1cm} & $0.2$ & $1.22$ & $1.22$ & $1.22$ & $1.23$ & $1.22$\\
& $0.1$ & $1.24$ & $1.25$ & $1.26$ & $1.26$ & $1.26$\\
& $0$ & $1.27$ & $1.29$ & $1.30$ & $1.31$ & $1.31$\\
\hline
\end{tabular}
\end{table}

We first discuss the predictions of our model at turnaround. The ratios between the different values of $R_\rmn t$ and $\Delta_\rmn t$ obtained from the SCM and from the SCM-L are listed in table \ref{tab:Rt} and in table \ref{tab:Deltat}. The SCM-L predicts smaller turnaround radii consistently with higher turnaround overdensities. This means that the presence of angular momentum slows down the infall motion of cluster components in its outskirts. In fact, a lower infall velocity implies that the turnaround region has to be shifted inwards, to higher-density regions. This effect is remarkable and is consistent with the results of CMM08 about the velocity profiles, obtained from the statistical analysis of simulated clusters. The ratios between the SCM and the SCM-L outcomes for  $R_\rmn t$ and $\Delta_\rmn t$, i.e. $R_\rmn{t,SCM}/R_{\rmn{t,SCM}-\rmn{L}}$ and $\Delta_\rmn{t,SCM}/\Delta_{\rmn{t,SCM}-\rmn{L}}$, are quite similar for different choices of $\Omega_{\rmn M,0}$ and $z_\rmn f$ (see tables \ref{tab:Rt} and \ref{tab:Deltat}).

In order to show the dependences of the obtained values of $\Delta_\rmn t$, $M_\rmn t/M_\rmn v$, and $r_\rmn t/r_\rmn v$ on $\Omega_{\rmn M,0}$ and $z_\rmn f$, we obtained the following analytical fittings, which reproduce the predictions of the models with an accuracy better than 3 per cent for $0.2\le\Omega_{\rmn M,0}\le 0.4$ and $0\le z_\rmn f\le 0.5$:
\be\label{eq:Deltat_fit}
\Delta_\rmn{t,fit}\equiv\left\{
\begin{array}{lll}
12.2\left(\f{\Omega_{\rmn M,0}}{0.3}\right)^{-0.66}(1+z_\rmn f)^{-1.91}&&\textrm{for SCM}\vspace{0.2cm}\\
15.8\left(\f{\Omega_{\rmn M,0}}{0.3}\right)^{-0.64}(1+z_\rmn f)^{-2.2}&&\textrm{for SCM-L}
\end{array}
\right.;
\ee
\be\label{eq:Mt_fit}
\left.\f{M_\rmn{t}}{M_\rmn{v}}\right|_\rmn{fit}=\left\{
\begin{array}{lll}1.90\left(\f{\Omega_{\rmn M,0}}{0.3}\right)^{0.145}(1+z_\rmn f)^{1.23}&&\textrm{for SCM}\vspace{0.2cm}\\
2.1\left(\f{\Omega_{\rmn M,0}}{0.3}\right)^{0.094}(1+z_\rmn f)^{1.45}&&\textrm{for SCM-L}
\end{array}
\right.;
\ee
\be\label{eq:rt_fit}
\left.\f{r_\rmn{t}}{r_\rmn{v}}\right|_\rmn{fit}=\left\{
\begin{array}{lll}
3.7\left(\f{\Omega_{\rmn M,0}}{0.3}\right)^{0.066}(1+z_\rmn f)^{1.05}&&\textrm{for SCM}\vspace{0.2cm}\\
3.6\left(\f{\Omega_{\rmn M,0}}{0.3}\right)^{0.043}(1+z_\rmn f)^{1.22}&&\textrm{for SCM-L}
\end{array}
\right..
\ee
Values obtained from equations (\ref{eq:Deltat_fit}), (\ref{eq:Mt_fit}), and (\ref{eq:rt_fit}) in the abovementioned ranges of $\Omega_{\rmn M,0}$ and $z_\rmn f$ are represented with a grey scale in figure (\ref{fig:Deltat_Mt_rt}) for both the SCM and the SCM-L (upper and lower panel, respectively). The difference between the two models mostly affects the values of $\Delta_\rmn t$, while it is less evident for the ratios $M_\rmn{t}/M_\rmn{v}|_\rmn{fit}$ and $r_\rmn{t}/r_\rmn{v}|_\rmn{fit}$. This indicates that the turnaround scale and the virialization scale are similarly affected by the introduction of the angular momentum term into the evolution equation. It means that the values of $M_\rmn{t}/M_\rmn{v}$ and $r_\rmn{t}/r_\rmn{v}$ obtained by CMM08 in the SCM scenario do not differ significantly from the values one should obtain in the SCM-L scenario.

Both SCM and SCM-L are in quite good agreement with the observations and the analysis of cosmological simulations for $\Omega_\rmn{M,0}=0.3$ and $z_\rmn f=0$. CMM08 evaluated a turnaround overdensity $\Delta_\rmn t=16_{-3}^{+4}$, taking into account cosmological simulations \citep{Borgal,Borgal4,Bivial}, which was only marginally consistent with the SCM scenario. This discrepancy  fosters the necessity of considering the angular momentum in the SCM. Our new SCM-L succeeds to be in quite good agreement also with \cite{RD} (who found a ratio between the turnaround mass and the virialization mass equal to about two\footnote{The original result is $M_\rmn t/M_{200}=2.18\pm 0.19$, where $M_{200}$ is the mass enclosed in a sphere with overdensity $\Delta=200$. Assuming that $M_\rmn v\simeq 1.1 M_{200}$, according to what stated by \cite{RD}, we obtain $M_\rmn t/M_\rmn v=2.0\pm 0.2$.}). 

\subsection{Profiles in the non-equilibrium region}\label{sec:prof}

We now consider the mass and velocity profiles predicted by of the SCM-L with $\Omega_{\rmn M,0}=0.3$ and $z_\rmn f=0$, coherent with the concordance $\Lambda$-CDM cosmology and the near cluster catalogues. To better appreciate our results, we compare our theoretical profiles with those obtained from a catalogue of 9631 member galaxies identified within 114 simulated clusters by \cite{Bivial} (see \citealt{Borgal} and  \citealt{Borgal4} for details on the cosmological simuation; the catalogue is described also in CMM08 and CMM10). 

\begin{figure}
\centering
\includegraphics[width=8.5cm]{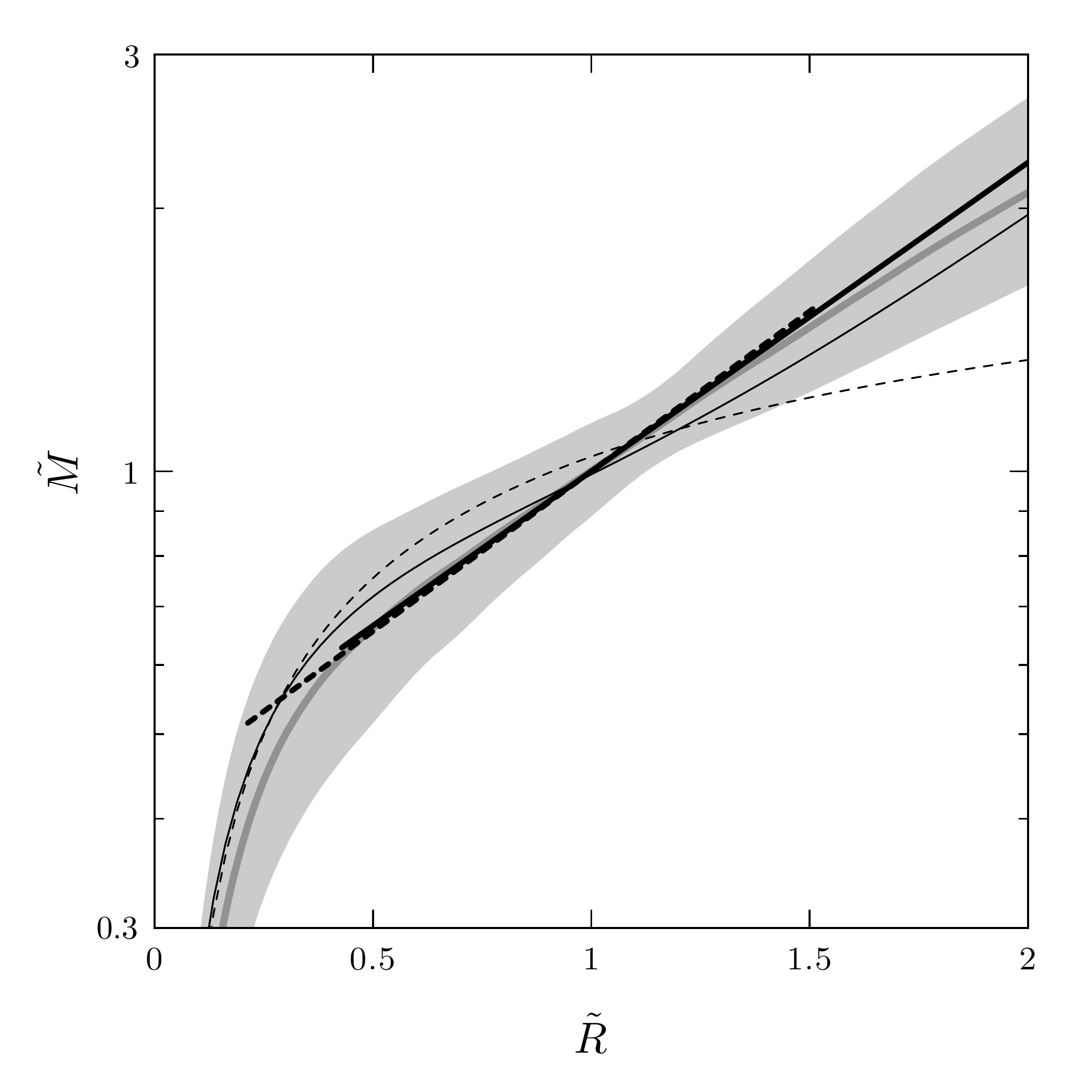}
\caption[]{Comparison between mass profiles. Bold dashed line: SCM prediction; bold solid line: SCM-L prediction; narrow dashed line: NFW profile \citep{NFW95,NFW96,NFW97}; narrow solid line: modified NFW profile \citep{Tal}. The grey solid line and the shaded area represent the mean profile and the 1$\sigma$ region as extracted from a catalogue of 9631 member galaxies \citep{Borgal,Bivial}.}
\label{fig:tildeM}
\end{figure}

The profiles of $\tilde M=M/M_\rmn t$ versus $\tilde R=R/R_\rmn t$ are shown in figure \ref{fig:tildeM}. The theoretical profiles predicted by SCM (bold dashed line) and by the SCM-L (bold solid line) are superimposed to the mean profile obtained from the distribution of the 9631 member galaxies (grey solid line); the 1$\sigma$ region of the distribution of cluster mass profiles is also shown (shaded region). Both the SCM and the SCM-L are in very good agreement with the mean simulated profile. 
Nevertheless, the SCM-L turns out to be slightly more accurate to describe the mass distribution beyond the turnaround region. 

The NFW mass profile by \cite{NFW95,NFW96,NFW97} is also plotted in figure \ref{fig:tildevparallel} (narrow dashed line). As already pointed out by CMM08, the NFW profile seems not to successfully describe the non-equilibrium region of galaxy cluster. We take into account also the modification of the standard NFW profile suggested by \cite{Tal} (narrow solid line). Evidently, this profile is in better agreement with the simulated cluster distribution and is fairly consistent with our SCM-L prediction; it evidences a significant convergence between \cite{Tal} and our approach.

\begin{figure}
\centering
\includegraphics[width=8.5cm]{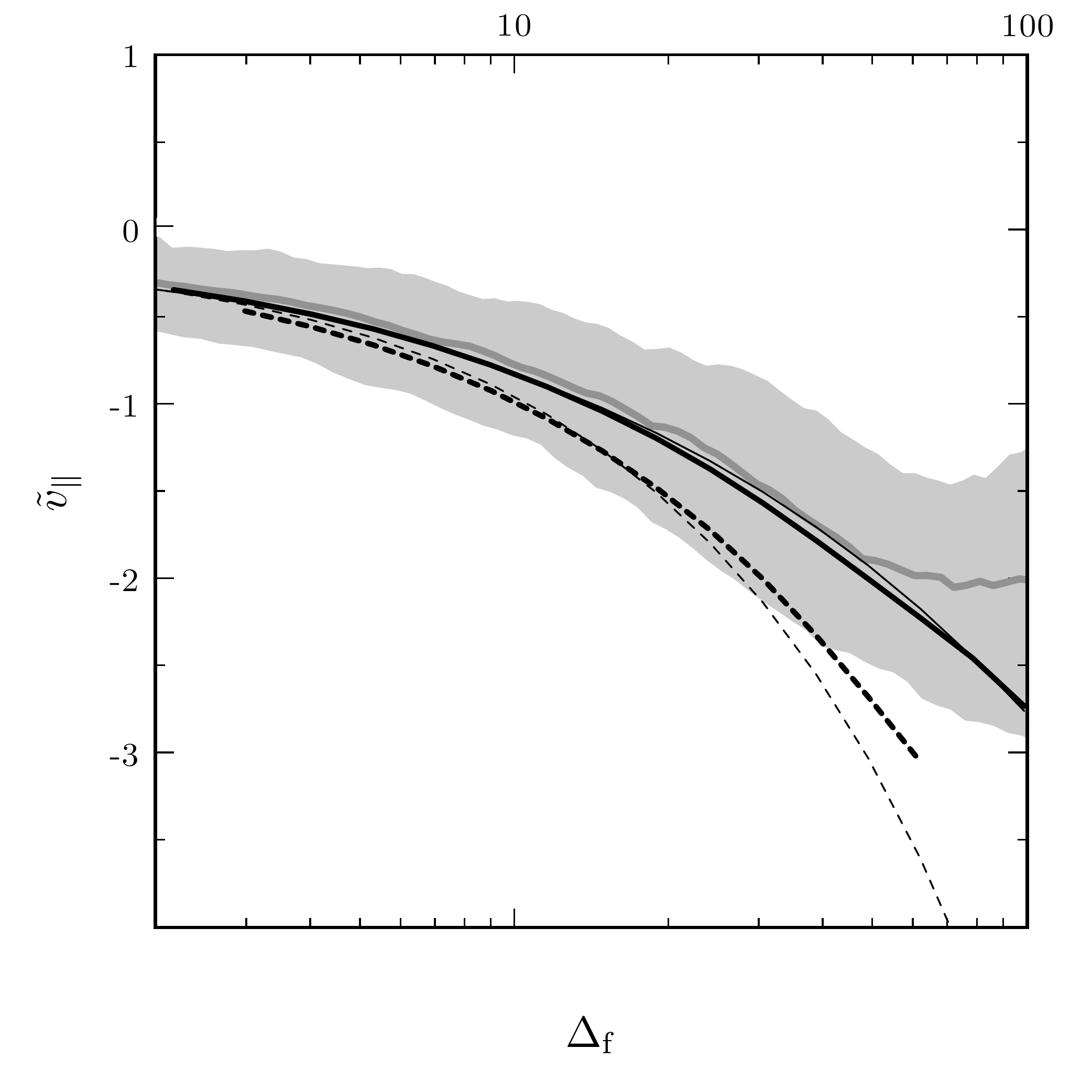}
\caption[]{Comparison between radial velocity profiles. Bold dashed line: SCM prediction; bold solid line: SCM-L prediction; narrow dashed line: Yahil approximation \citep{Y}; narrow solid line: Meiksin approximation \citep{VD}. The grey solid line and the shaded area represent the mean profile and the 1$\sigma$ region as extracted from a catalogue of 9631 member galaxies \citep{Borgal,Borgal4,Bivial}.}
\label{fig:tildevparallel}
\end{figure}

The radial velocity profiles of the cluster components are shown in figure \ref{fig:tildevparallel}. Once again, the theoretical profiles predicted by SCM (bold dashed line) and by the SCM-L (bold solid line) are superimposed to the mean profile obtained from the distribution of the 9631 member galaxies (grey solid line); the 1$\sigma$ region of the distribution of cluster mass profiles is also shown (shaded region). Here the difference between the SCM and the SCM-L stands out clearly, in fact the two models predict different values of $\Delta_\rmn f$ at turnaround. In particular, the SCM gives lower values of $\Delta_\rmn t$ than the SCM-L does, with a consequent steeper velocity profile in the SCM scenario. The SCM systematically overestimates the amount of radial velocity in the whole overdensity range from the extreme outskirts of clusters to the virialization core. This is not the case of the SCM-L scenario, due to the presence of the angular momentum term, which has been shown to be responsible for slowing down the infall motion in the non-equilibrium regions. The abovementioned agreement between the SCM-L prediction and the simulated mean profile gives a strong support to the hypothesis that angular momentum, due to the motion of galaxy groups in cluster outskirts, has to be taken into account. The agreement also justifies \emph{a posteriori} our assumption that the dominant contribution to angular momentum comes from scales of the order of $\sim$$10 h^{-1}$ Mpc (see section \ref{sec:ang_mom}).

Two different literature approximations for the radial velocity profile are also shown in figure \ref{fig:tildevparallel}, namely the Yahil approximation $\tilde v_\rmn{Y}$ (\citealt{Y}; narrow dashed line),
\be
\tilde v_\rmn Y(\Delta)=\f{\Delta-1}{3\Delta^{1/4}},
\ee
and the Meiksin approximation $\tilde v_\rmn{M}$ (\citealt{VD}; narrow solid line),
\be
\tilde v_\rmn M(\Delta)=\f{\sqrt{3}}{3}\f{\Delta-1}{(\Delta+2)^{1/2}}.
\ee
The Yahil approximation is in good agreement with the profile predicted by the standard SCM for $\Delta\lesssim 30$, while the Meiksin approximation is nearly coincident with our new profile predicted by the SCM-L, and thus it seems to be the best choice. It confirms what was found by CMM08.

\begin{figure}
\centering
\includegraphics[width=8.5cm]{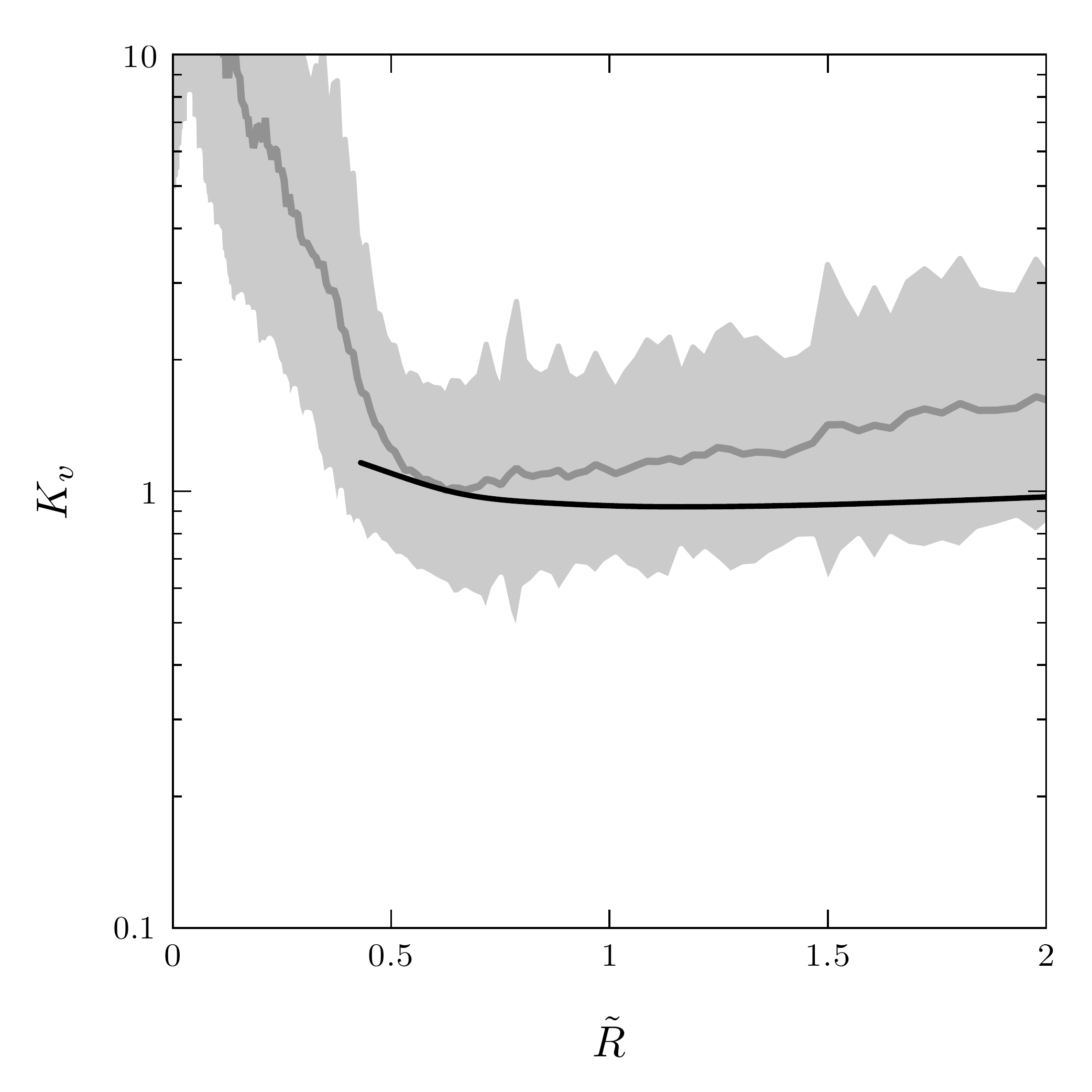}
\caption[]{Ratio between tangential and radial velocity. Bold solid line: SCM-L prediction. The grey solid line and the shaded area represent the mean profile and the 1$\sigma$ region as extracted from a catalogue of 9631 member galaxies \citep{Borgal,Bivial}.}
\label{fig:Kv}
\end{figure}

We also computed the ratio between the tangential velocity and the radial velocity predicted by the SCM-L, $K_v\equiv v_\perp/v_\parallel$, as a function of the radial coordinate $\tilde R = R/R_\rmn t$ (where $R_\rmn t$ is the turnaround radius). The profile obtained from this latter function is shown in figure \ref{fig:Kv}, superimposed to the mean profile obtained from the distribution of the 9631 member galaxies (grey solid line); the 1$\sigma$ region of the distribution is also shown (shaded region). The quantity $K_v$ is evidenced by CMM10 to be crucial to relate the mean observed line-of-sight velocity of member galaxies in clusters to their mean infall velocity (towards the cluster centre). The knowledge of the mean infall velocity profile is crucial to estimate the mass profile (see CMM10 for the details). 

The SCM-L prediction of $K_v$ agrees within the 1$\sigma$ uncertainty with the mean profile traced by the simulated galaxies, and it is nearly constant along the radial direction, where the value of $K_v$ is about $0.95$ in the turnaround region. $K_v=0.95$ is the fiducial value adopted by CMM10 to identify the `fair region' in the redshift distribution of galaxies, i.e. the region where the galaxies are shown to better trace the infall motion, and subsequently to better obtain the cluster mass profile.


\section{Conclusions}\label{sec:concl}

Our new formulation of the Spherical Collapse Model (SCM-L) takes into account the presence of angular momentum associated with the motion of galaxy groups infalling towards the centre of galaxy clusters. The angular momentum (estimated from the power spectrum of density perturbations) is responsible for an additional term in the dynamical equation which is useful to describe the evolution of the clusters in the non-equilibrium region. Once the evolution equation is solved, our SCM-L provides a complete description of the dynamical state of the perturbations at any given redshift $z$ as a function of the adopted values of matter density function $\Omega_0$.

We run our model with different values $\Omega_0$ between $0.2$ and $0.4$ and redshift $z$ between $0$ and $0.5$. In each case, we computed the turnaround overdensity $\Delta_\rmn t$ and the ratio between the turnaround scale and the virialization scale. 

Our SCM-L can be also used to predict the profiles of several strategic dynamical quantities (radial and tangential velocities of member galaxies, and total cluster mass). The agreement between our SCM-L and the simulations is always remarkable in the whole non-equilibrium region (spanning approximately from $0.5$ to $2$ turnaround radii $r_\rmn t$), which is the region of interest of the present paper. 

The non-equilibrium region is the natural scenario where to study the infall in galaxy clusters and the accretion phenomena present in these objects. The method described in the present paper can be also generalized to alternative cosmological models. 

Our results corroborate our previous estimates (CMM08); moreover, they are in very good agreement with the analyses of recent observations \citep{DG} and of simulated clusters (see, e.g., CMM10).
\section*{ACKNOWLEDGMENTS}

We wish to thank Stefano Borgani for making available to us the simulated data, Andrea Biviano and Marisa Girardi for providing the simulated galaxy catalogue, and all of them for the useful discussions and insightful advices. We wish to thank the anonymous Referee too, for the discussion and the useful suggestions.


\begin{figure}
\centering
\includegraphics[width=8.5cm]{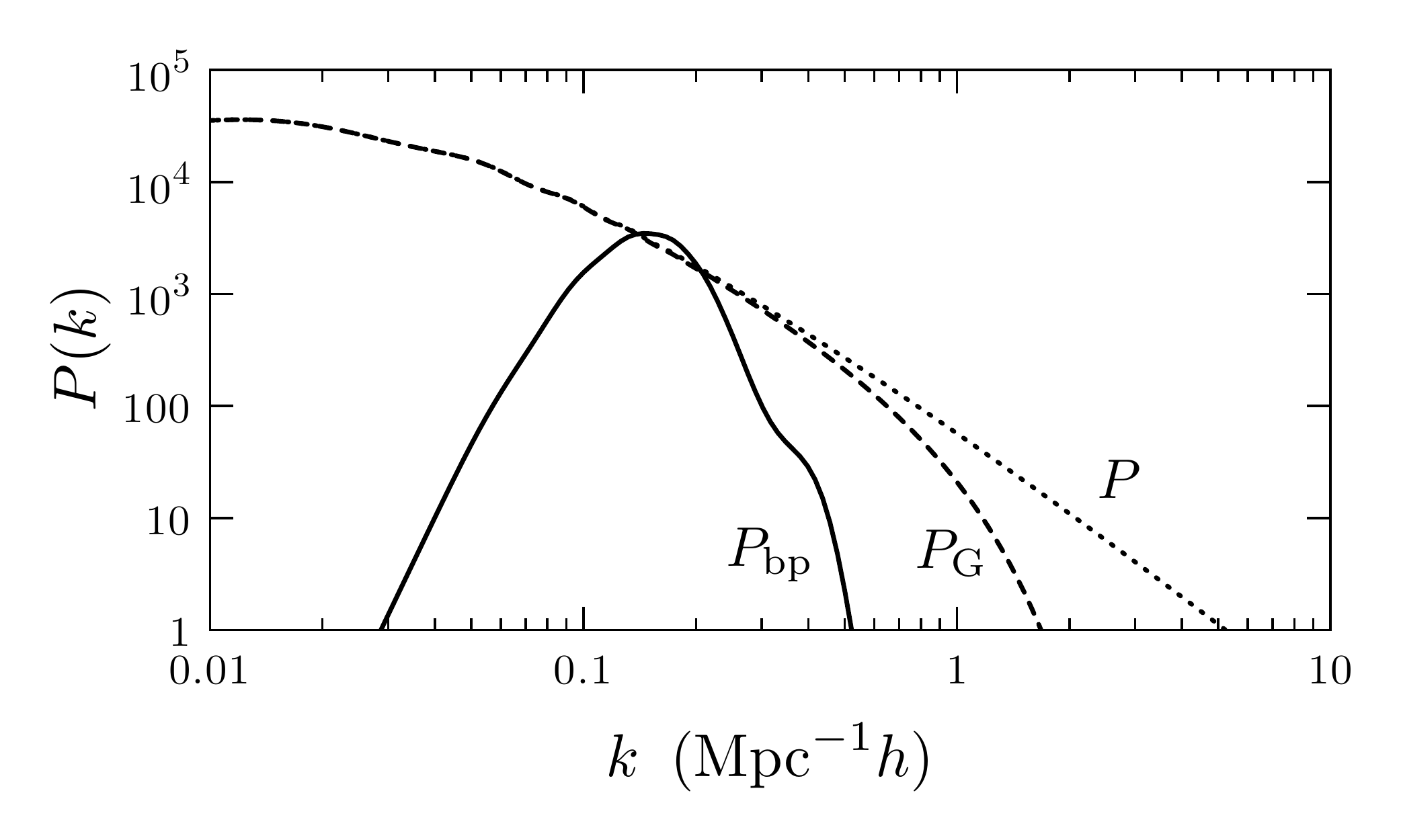}
\caption[]{Power spectrum (linear) of the perturbation obtained with different filters. The dotted line represents the original power spectrum, $P(k)$, the dashed line represents the Gaussian-filtered power spectrum, $P_\rmn G(k)$, while the solid line represents the power spectrum filtered at the group scale, $P_\rmn{bp}(k)$.}
\label{fig:P}
\end{figure}

\appendix

\section{Primordial overdensity of the perturbations}\label{app:delta_i}

This appendix provides the recipe for computing the initial overdensity profile of a spherically symmetric density perturbation from the power spectrum. We use the results of \cite{BBKS}, \cite{Lal}, and \cite{LL}; see also \cite{CMM08} for an example of the same approach. 

The overdensity profile $\delta_\rmn i$ (within $r_\rmn i$) is obtained from the statistic of peaks in the density field described by the power spectrum. To identify the progenitors at the cluster scale, we apply a gaussian filter with radius $r_\rmn G=1h^{-1}\rmn{Mpc}$ to isolate the contribution of the galaxy cluster scale: 
\be\label{eq:gauss_filt}
P_\rmn G(k)\equiv P(k)e^{-(r_\rmn G k)^2/2};
\ee
$P_\rmn G(k)$ is represented as a dashed line in figure \ref{fig:P}; the original power spectrum $P(k)$ is also represented for comparison (dotted line). We denote with $\sigma_l^2$ the $(l+1)$-th even moment of $P_\rmn G(k)$:
\be
\sigma_l\equiv\sqrt{\f{1}{2\pi^2}\int_0^\infty{P_\rmn G(k)k^{2(l+1)}\rmn dk}},
\ee
If $\nu\sigma_0$ is the expected height of a peak, we can express the differential number density of peaks as a function of $\nu$:
\be
\mathcal N(\nu)=\f{1}{(2\pi)^2r_*^3}e^{-\nu^2/2}\Xi(\gamma,\gamma\nu),
\ee
\be
\quad\gamma\equiv\f{\sigma_1^2}{\sigma_0\sigma_2},\quad r_*\equiv\sqrt 3\f{\sigma_1}{\sigma_2},
\ee
\be
\quad\Xi(\gamma,w)\equiv\f{w^3+3\gamma^2 w+\big[A(\gamma)w^2+B(\gamma)\big]e^{-C(\gamma)w^2}}{1+D(\gamma)e^{-E(\gamma)w}},
\ee
\be
\qquad A(\gamma)\equiv\f{432}{\sqrt{10\pi}(9-5\gamma^2)^{5/2}},
\ee
\be
\qquad B(\gamma)\equiv 1.84+1.13(1-\gamma^2)^{5.72},
\ee
\be
\qquad C(\gamma)\equiv\f{5}{2(9-5\gamma^2)},
\ee
\be
\qquad D(\gamma)\equiv 8.91+1.27e^{6.51\gamma^2},
\ee
\be
\qquad E(\gamma)\equiv 2.58e^{1.05\gamma^2}.
\ee
The linear approximation for the present-day overdensity profile is consequently obtained as
\begin{eqnarray}\label{eq:delta_ti}
\delta_{\rmn{lin},0}(r)&=&\f{3}{2\pi^2\sigma_0 r}\int_0^\infty k j_1(kr) P_{\rmn f,1}(k)\times\nonumber\\
&&\times\left[\f{\bar\nu-\gamma^2\bar\nu-\gamma\bar\theta}{1-\gamma^2}+\f{\bar\theta r_*^2}{3\gamma(1-\gamma^2)}k^2\right]dk.
\end{eqnarray}
\be
\quad\bar\nu\equiv\f{\int_{\nu_\rmn{thr}}^\infty{\nu\mathcal N(\nu)\rmn d\nu}}{\int_{\nu_\rmn{thr}}^\infty{\mathcal N(\nu)\rmn d\nu}},\quad\bar\theta\equiv\f{\int_{\nu_\rmn{thr}}^\infty{\theta(\nu)\mathcal N(\nu)\rmn d\nu}}{\int_{\nu_\rmn{thr}}^\infty{\mathcal N(\nu)\rmn d\nu}},
\ee
\be
\qquad\theta(\nu)\equiv\f{3(1-\gamma^2)+(1.216-0.9\gamma^4)e^{-(\gamma/2)(\gamma\nu/2)^2}}{\sqrt{3(1-\gamma^2)+0.45+\left(\f{\gamma\nu}{2}\right)^2}+\f{\gamma\nu}{2}}.
\ee

We assume that at the initial time the perturbation is not yet uncoupled from the Hubble flow, and that the initial overdensity is thus well predicted by the linear approximation: $\delta_\rmn i=\delta_\rmn{lin,i}$. This requirement is generally satisfied at high redshift: $z_\rmn i\simeq 1000$. To compute the initial profile $\delta_\rmn{lin,i}(r)$, we simply revert back the evolved profile $\delta_{lin,0}(r)$ using the growing-mode linear solution of the SCM:
\be
\delta_\rmn{lin,i}(r)=\delta_{\rmn{lin},0}(r)\f{D_+(z_i)}{D_{+}(0)},
\ee
\begin{eqnarray}
\quad D_+(z)&\equiv&\f{5\Omega_\rmn M(z)}{2(1+z)}\Big[\Omega_\rmn M(z)^{4/7}-\Omega_\Lambda(z)+\nonumber\\
&&+\left(1+\f{\Omega_\rmn M(z)}{2}\right)\left(1+\f{\Omega_\Lambda(z)}{70}\right)\Big]^{-1}.
\end{eqnarray}

\section{Friedman equation with angular momentum}\label{app:fried}

This appendix describe the technical details concerning the evolution of a spherically symmetric density perturbation with angular momentum. We first discuss how to obtain the angular momentum from the power spectrum of perturbations and then detail the routine to numerically solve the Friedman equation (\ref{eq:fried}). 

The angular momentum can be computed from the peculiar velocity field obtained from the power spectrum. To isolate the angular momentum contribution of galaxy groups, we adopt the following band-pass filter, using a scale radius $r_\rmn{cl}$ for the cluster scale and a scale radius $r_\rmn{gr}$:
\begin{eqnarray}\label{eq:bp_filt}
P_\rmn{bp}(k)&\equiv& P(k)\left[1-\f{\sin(kr_\rmn{cl})}{kr_\rmn{cl}}\right]^2\times\nonumber\\
&\times&\left\{\f{3\big[\sin(kr_\rmn{gr})-kr_\rmn{gr}\cos(kr_\rmn{gr})\big]}{(kr_\rmn{gr})^3}\right\}.
\end{eqnarray}
$P_\rmn{bp}(k)$ is represented as a dashed line in figure \ref{fig:P}; the original power spectrum $P(k)$ is also represented for comparison (dotted line). The radii $r_\rmn{cl}$ and $r_\rmn{gr}$ are computed as the mean separation between neighbouring clusters and neighbouring groups, respectively, defined as functions of the cluster number density $n_\rmn{cl}$ and of the group number density $n_\rmn{gr}$:
\be
r_\rmn{cl}=\left(\f{3}{4\pi n_\rmn{cl}}\right)^{1/3},\quad r_\rmn{gr}=\left(\f{3}{4\pi n_\rmn{gr}}\right)^{1/3}.
\ee
The cluster number density is extracted from the large cosmological simulation of \cite{Borgal,Borgal4} (see also \citealt{Bivial}; CMM08; CMM10). The estimate for galaxy groups is taken from  \cite{GeG}, and corresponds to the number density of objects with mass larger than $9\times 10^{12} h^{-1} M_ {\odot}$ (\ldots)
\be
n_\rmn{cl}\simeq 1.6\times 10^{-5}h^3\,\rmn{Mpc}^{-3},\quad r_\rmn{cl}=24.6 h^{-1}\,\rmn{Mpc};
\ee
\be
n_\rmn{cl}\simeq 1.0\times 10^{-3}h^3\,\rmn{Mpc}^{-3},\quad r_\rmn{cl}=6.20 h^{-1}\,\rmn{Mpc}.
\ee
The peculiar velocity field is obtained through the continuity relation between $\delta$ and $v$.
If the motion of the group is only due to large scale perturbations, $\delta$ is well predicted by the linear approximation. We obtain in this case
\begin{eqnarray}\label{eq:v_field}
\langle v^2\rangle&=&\f{H^2 f_\delta^2 a^2}{2\pi^2}\int_0^\infty{P_\rmn{bp}(k)\rmn dk}\nonumber\\
&=&\f{H^2 f_\delta^2 D_+}{(1+z)^2}\int_0^\infty{\f{k^3 P_\rmn{bp}(k)}{2\pi^2}\f{a_{0}^2}{k^2}\f{\rmn dk}{k}};
\end{eqnarray}
\be
\quad f_\delta= \f{F(\delta)}{\delta}\equiv\f{\rmn d\ln D_+}{d\ln a_0}.
\ee 
The ortogonal velocity $v_\perp$ is obtained as the projection of $v$ onto a plane (cf. section \ref{sec:ang_mom}). Under the isotropy assumption, this gives $v_\perp=\sqrt{2\langle v^2\rangle/3}$. From  equation (\ref{eq:v_field}) 
we finally obtain
\be\label{eq:L_ev}
C(z)=a v_\perp =H_0 a_0KQ(z),
\ee
\be
\quad K\equiv\sqrt{\f{2}{3}\int_0^\infty{\f{k^3 P_\rmn{bp}(k)}{2\pi^2}\f{a_{0}^2}{k^2}\f{\rmn dk}{k}}},
\ee
\be
\quad Q(z)\equiv\f{E(z)f_\delta(z)D_+(z)}{(1+z)^2}.
\ee

The knowledge of $\mathcal L$ allows the estimation of $\bar\mathcal L$ from equation (\ref{eq:L_comp}).
Equation (\ref{eq:fried}) and equation (\ref{eq:L_comp}) have to be solved together, since both the radius $r$ and the angular momentum $\bar\mathcal L$ change in time. After rearranging, we obtain
\begin{eqnarray}\label{eq:system_fried}
\ddot R(\tau)&=&-2\f{\dot a(\tau)}{a(\tau)}\dot R(\tau)+\Big[(1-\Omega_{\rmn M,0})-\f{\ddot a(\tau)}{a(\tau)}\Big]R(\tau)+\nonumber\\
&-&\f{\Omega_{\rmn M,0}\big[1+\delta_\rmn i(r_\rmn i)\big]}{2 R(\tau)^2}\left[\f{a(\tau)}{a_0}\right]^{-3}\!\!+\f{\bar\mathcal L(\tau)^2}{R(\tau)^3}\left[\f{a(\tau)}{a_0}\right]^{-4}\!\!;
\end{eqnarray}
\be\label{eq:system_L}
\dot{\bar\mathcal L}(\tau)=K R\dot Q(\tau);
\ee
\begin{eqnarray}
\quad\tau\equiv H_0 t=\f{2}{3\sqrt{1-\Omega_{\rmn M,0}}}\rmn{asinh}\left[\sqrt{\f{1-\Omega_{\rmn M,0}}{\Omega_{\rmn M,0}}}(1+z)^{-3/2}\right]\!\!,\nonumber\\
\end{eqnarray}
\be
\quad R(\tau)\equiv\f{r(\tau)}{a(\tau)};
\ee
where the dots denote first- and second-order derivatives with respect to $\tau$. The boundary counditions needed to solve the system (\ref{eq:system_fried})-(\ref{eq:system_L}) are set by assuming that at the initial time the perturbation is not yet uncoupled from the Hubble flow, i.e. $r_i=a_i$, but the peculiar radial velocity is given by the linear reltion. This requirement is satisfied at high redshift: $z_\rmn i\simeq 1000$.
Consequently, 
\be
R_\rmn i=1,
\ee
\be
\dot R_\rmn i=-\f{1}{3}f_{\delta,\rmn i}\delta_\rmn i\Omega_{\rmn M,0}^{1/2}(1+z_\rmn i)^{3/2},
\ee
\be
\bar\mathcal L_\rmn i=KQ_\rmn i.
\ee

Once solved, the system (\ref{eq:system_fried})-(\ref{eq:system_L}) provides the values of $R_\rmn f$, $\dot R_\rmn f$ and $\bar\mathcal L_\rmn f$ for any given value of $\tau_\rmn f>\tau_\rmn i$. All the relevant physical properties of the perturbation, including the infall velocity $v_{r,\rmn f}$, are consequently obtained as follows ($r_i$ is given in comoving units):
\be
r_\rmn f=R_\rmn f r_i;
\ee
\be
\delta_\rmn f=(1+\delta_\rmn i)R_\rmn f^{-3}-1;
\ee
\be
v_{\parallel,\rmn f}=\f{\dot R_\rmn f}{R_\rmn f}H_0 R_\rmn f r_i;
\ee
\be
v_{\perp,\rmn f}=\f{\bar\mathcal L_\rmn f}{R_\rmn f^2}H_0 R_\rmn f r_i.
\ee
}

\end{document}